\RequirePackage{lineno} 
\documentclass[singlecolumn,floatfix,amsmath,amssymb,citeautoscript,showpacs]{revtex4-1}

\usepackage{amsmath}
\usepackage{fancyhdr}
\usepackage{afterpage}

\usepackage{graphicx}
\usepackage{dcolumn}
\usepackage{bm}
\usepackage{amsmath}
\usepackage{amssymb}
\usepackage{listings}
\usepackage{verbatim}
\usepackage{cancel}

\newcommand*\patchAmsMathEnvironmentForLineno[1]{%
  \expandafter\let\csname old#1\expandafter\endcsname\csname #1\endcsname
  \expandafter\let\csname oldend#1\expandafter\endcsname\csname end#1\endcsname
  \renewenvironment{#1}%
     {\linenomath\csname old#1\endcsname}%
     {\csname oldend#1\endcsname\endlinenomath}}%
\newcommand*\patchBothAmsMathEnvironmentsForLineno[1]{%
  \patchAmsMathEnvironmentForLineno{#1}%
  \patchAmsMathEnvironmentForLineno{#1*}}%
\AtBeginDocument{%
\patchBothAmsMathEnvironmentsForLineno{equation}%
\patchBothAmsMathEnvironmentsForLineno{align}%
\patchBothAmsMathEnvironmentsForLineno{flalign}%
\patchBothAmsMathEnvironmentsForLineno{alignat}%
\patchBothAmsMathEnvironmentsForLineno{gather}%
\patchBothAmsMathEnvironmentsForLineno{multline}%
}

\newcommand{\ba}{\begin{align}}
\newcommand{\ea}{\end{align}}

\newcommand{\bea}{\begin{eqnarray}}
\newcommand{\eea}{\end{eqnarray}}
\def\<#1\>{\begin{align}#1\end{align}}

  \makeatletter
  \renewcommand{\Large}{\@setfontsize\Large{13.5}{15.5}}
  \makeatother

\begin{document}
\author{Karthik Sasihithlu}%
\affiliation{%
 Department of Energy Science and Engineering, Indian Institute of Technology Bombay, Mumbai 400076, India
}%
\email{ksasihithlu@iitb.ac.in}

\date{\today}



\title{A coupled harmonic oscillator model to describe the near-field radiative heat transfer
between nanoparticles and planar surfaces}

\begin{abstract}
When two objects made of a material which supports surface modes  are brought in close proximity to each other such that the vacuum gap between them is less than the thermal wavelength of radiation, then the coupling between the surface modes provides an important channel for the heat transfer to occur which is different from that mediated by long range propagating electromagnetic waves. Indeed, the heat transfer then exceeds Planck's blackbody limit by several orders of magnitude, and consequently has been used for several energy applications such as near-field thermophotovoltaic systems. This near-field radiative heat exchange has been traditionally and successfully described using fluctuational electrodynamics principles.  Here, we describe an alternate coupled harmonic oscillator model approach which can be used to model the coupling between surface modes and hence the resultant near-field heat transfer.  We apply this theory to estimate the near-field heat transfer for the configurations of two metallic nanoparticles and two planar metal surfaces and compare the result with predictions from fluctuational electrodynamics theory. 
\end{abstract}

\maketitle

\section{Introduction}
\label{section2}

The discovery that a peak in the density of states at certain frequencies where surface modes can be thermally excited will  result in a significant enhancement in heat transfer when objects are brought in close proximity to each other has lead to development of applications relevant to energy systems such as near-field thermophotovoltaics \cite{narayanaswamy03a, dimatteo2001enhanced, laroche06b} and thermal rectification \cite{otey2010thermal}.%
This enhancement in heat transfer which can be several orders higher than that exchanged by objects when separated by large distances ($d \gg \lambda_T$ where $\lambda_T$ is the thermal wavelength of radiation) has been successfully explained using the theory of fluctuational electrodynamics developed by Rytov \cite{rytov59a} and first applied for analysing heat transfer between closely spaced bodies by Polder and Van Hove \cite{polder71}.
The theoretical framework of this procedure relies on the introduction of external microscopic thermal fluctuating currents or dipole moments whose correlation functions are related to the dielectric properties of the material via the fluctuation-dissipation theorem \cite{loomis94, pendry1999radiative, joulain05a}.
%
While the predictions from this theory has indeed been experimentally verified over the years for several geometries including that for flat surfaces \cite{song2016radiative}, STM tip over substrate  \cite{kim2015radiative, kloppstech2017giant} and sphere over a flat substrate  \cite{rousseau2009radiative, shen2009surface}, a main assumption in this theory is the prevalence of local thermodynamic equilibrium in the objects which restricts the cases to be analysed to stationary or quasi-stationary cases only.
\\
Recently, an alternate approach which uses the coupled harmonic oscillator theory to model the coupling between surface polaritons has been proposed which enables us to  analyse the near-field  heat transfer for both dynamic situations valid for time scales less than the relaxation time scales of surface polariton excitations, as well as in the steady-state \cite{biehs2013dynamical, barton2015classical, sasihithludynamic}.   In this work we adopt this approach to arrive at expression for the steady-state near-field heat transfer for the configurations of two nanoparticles and also for two planar systems and compare the expressions  with that derived using flucutational electrodynamics. The approach we use in this study differs from that explained in Ref. 
  The advantage of this model is the simplicity and the generality that this offers vis-a-vis the flucutational electrodynamics theory.  By modelling the coupling of surface modes using the harmonic oscillator model, the parameters thus derived will be useful to analyze not only heat transfer but also other phenomena  where the resonant excitation of surface modes plays an important role such as van der Waals forces,  decreased lifetime of molecules close to surfaces  \cite{chance1978molecular}, and surface enhanced Raman scattering \cite{le2008principles}, while also giving us additional insight into the mechanism and the strength of coupling between surface modes.

The paper is arranged as follows:  in Section \ref{section3} the theory of steady state heat transfer between coupled harmonic oscillators is developed and key results are highlighted.  In Section \ref{nano} this theory is applied to predict heat exchange between nanostructures and compared with the predictions of fluctuational electrodyanmics. In Section \ref{planar} we apply this theory to predict the near-field heat exchange between planar surfaces whose dielectric properties are given by the Drude model.

 \section{Heat transfer between coupled harmonic oscillators}
\label{section3}

  \begin{figure}[h]
\includegraphics[scale=0.45]{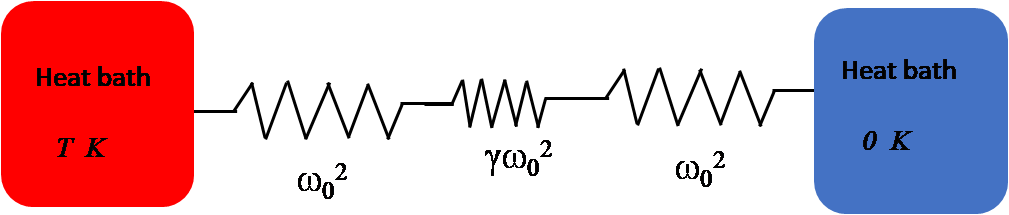}
\caption{The coupled harmonic oscillator system which is analysed in Section \ref{section3}. The oscillators with spring constant $\omega_0^2$ are connected to two separate heat baths each maintained at a different temperature. }
\label{FigHO}
\end{figure}

We model the surface modes between interacting systems as harmonic oscillators which are in contact with two separate heat baths maintained at constant temperatures as shown in Fig.  Since the oscillators are in contact with heat bath they are can be modelled as force driven.  While the heat flux between the two heat reservoirs for such a system has been well studied \cite{dorofeyev2013coupled, ghesquiere2013entanglement, biehs2013dynamical, barton2015classical}  here we provide an alternate derivation based on Green's function theory to arrive at the expression for heat transfer for such a system. A similar derivation has been given in Ref. \cite{pendry2016phonon}  in the context of modelling the heat transfer due to coupling of Rayleigh modes in solids. 
We begin with the equation of motion for such a system which in the frequency $\omega$ space can be written as:
\[
\omega^2 \hat{x}_1(\omega)= \omega_0^2 \hat{x}_1(\omega) - 2 i \omega \delta \hat{x}_1(\omega) + F(\omega, T)
 \]
where $F(\omega, T)$ is the forcing function whose spectral density $S_F(\omega,T)$ is as yet unknown, and $\delta$ is the half-line width. 
The Green's function for such a system defined such that $\hat{x}_1(\omega) =G(\omega)F(\omega, T)$  is given as:
\[
G(\omega) = \dfrac{1}{\omega^2 - \omega_0^2 + 2 i \omega \delta}
\]
Relating the density of states  $\rho(\omega)$ to the imaginary part of the Green's function, and recognising that it peaks around $\omega \approx \omega_0$, gives:
\[
\rho(\omega) = \dfrac{4}{\pi}\dfrac{\omega_0^2 \delta}{(\omega^2 - \omega_0^2)^2 + 4 \omega_0^2 \delta^2}
\]
At equilibrium at temperature $T$ we have the total energy contained in the harmonic oscillator, $E$, to be:
\begin{equation}
E = \int_0^\infty \rho(\omega)  \dfrac{\hbar \omega \, d\omega}{e^{\hbar \omega/(k_B T)} - 1}
\label{energy1}
\end{equation}
However, the energy contained in the oscillator is also given by: 
\begin{equation}
E = \langle \omega_0^2 x^2 (t)\rangle = \omega_0^2 \frac{\pi}{\Theta} \int_{-\infty}^{\infty} |x(\omega)|^2 \, d\omega = 2 \omega_0^2  \int_{0}^{\infty} |G(\omega)|^2 \, S_F(\omega, T) d\omega
\label{energy2}
\end{equation}
where, $\Theta$ is a large time interval and we make use of the definition of spectral density  $S_F(\omega, T) = (\pi/\Theta) |F(\omega, T)|^2$ from Ref. \cite{reif2009fundamentals}. 
%
%
%
%
%
%
%
%
%
Comparing Eq. \ref{energy1} and Eq. \ref{energy2} we obtain:
\begin{equation}
S_F(\omega, T) = \frac{2\delta}{\pi} \frac{\hbar \omega}{e^{\hbar \omega/k_B T} - 1}
\label{forcing}
\end{equation}
Now consider system of  two coupled oscillators with coupling constant $\gamma \omega^2_0$ such that one of the oscillators is in contact with a heat reservoir at temperature $T$ and the other at zero kelvin. The equations of motion can be written as:
\begin{eqnarray}
\omega^2 \hat{x}_1(\omega) &=& \omega_0^2 \hat{x}_1(\omega) - 2i \omega \delta \hat{x}_1(\omega) +  \gamma \omega_0^2 \hat{x}_2(\omega)+ F(\omega, T)  \nonumber \\
\omega^2 \hat{x}_2(\omega) &=& \omega_0^2 \hat{x}_2(\omega) -2 i \omega \delta \hat{x}_2(\omega) + \gamma \omega_0^2 \hat{x}_1(\omega)
\label{eqofmotion}
\end{eqnarray}
The eigenfrequencies of this coupled system in the absence of the forcing function and for $\Gamma/\omega_0 \ll 1$  is given as:
\begin{equation}
\omega^2_\pm = \omega_0^2 (1 \pm \gamma ) - i  \omega_0 \delta
\label{eqeigenHO}
\end{equation}
with the corresponding normalized eigenvectors given by: $ \frac{1}{\sqrt{2}}
\left[\begin{matrix} +1 \\ +1 \end{matrix} \right]$ and $ \frac{1}{\sqrt{2}}
\left[\begin{matrix} +1 \\ -1 \end{matrix} \right]$.
The new Green's function can be built from the eigenvectors as follows:
\begin{eqnarray}
\begin{split}
\bar{\bar{G}}(\omega) = \frac{1}{2}
\left[\begin{matrix} +1 \\ +1 \end{matrix} \right] \left[\begin{matrix} +1, +1  \end{matrix} \right] \dfrac{1}{\omega^2 - \omega_0^2 (1+ \gamma)+i \omega_0 \delta} +\\ \frac{1}{2}
\left[\begin{matrix} +1 \\ -1 \end{matrix} \right] \left[\begin{matrix} +1, -1  \end{matrix} \right] \dfrac{1}{\omega^2 - \omega_0^2 (1- \gamma) +i \omega_0 \delta}
\end{split}
\end{eqnarray}
Using this form of the Green's function we can write the response to be of the form:
\begin{eqnarray}
\left[\begin{matrix} \hat{x}_1(\omega) \\ \hat{x}_2 (\omega) \end{matrix} \right] = \dfrac{F(\omega, T)}{2} \left( \left[\begin{matrix} 1 \\ 1 \end{matrix}  \right] \dfrac{1}{\omega^2 - \omega_0^2 (1+ \gamma)+i \omega_0 \delta}  \right. \\ \left.  +\left[\begin{matrix} 1 \\ -1 \end{matrix}  \right] \dfrac{1}{\omega^2 - \omega_0^2 (1- \gamma)+i \omega_0 \delta} \right)
\end{eqnarray}
%
%
%
%
%
%
In steady state the rate of energy transfer to the second oscillator via coupling with the first oscillator is equal to the decay in the second oscillator via damping. This should be equal to the heat transfer to the sink. Thus we can write the expression for the heat transfer in the system shown in Fig.  as:
\begin{equation}
P = - \omega_0^2 \frac{d}{dt} \langle x_2^2 (t) \rangle = 2 \delta \omega_0^2 \langle x_2^2 (t) \rangle \label{damping}
\end{equation}
Following the procedure similar to that in Eq. \ref{energy2},  Eq. \ref{damping} further reduces to:
\begin{equation}
P = \delta \omega_0^2 \int_0^\infty S_F(\omega, T) \left|\frac{1}{(\omega^2 - \omega_0^2(1+\gamma) + i  \omega_0 \delta)} - \frac{1}{(\omega^2 - \omega_0^2 (1- \gamma)+i \omega_0 \delta)} \right|^2
\label{P1}
\end{equation}
%
%
%
%
%
%
%
%
%
%
Substituting the expression for $S_F(\omega, T)$ from Eq. \ref{forcing} and evaluating the integral while recognising that the integral is sharply peaked around $\omega \approx \omega_0$, we get:
 \begin{equation}
P = \frac{\hbar \omega_0}{e^{\hbar \omega_0/(k_B T) }-1} \dfrac{  \omega_0^2 \gamma^2 \delta}{(\omega_0^2\gamma^2 +  \delta^2)}
\label{eqHO}
\end{equation}
This expression matches the form for the steady-state heat transfer for this system derived by other authors using Master equation approach \cite{biehs2013dynamical} and Langevin theory \cite{ barton2015classical}.  

\section{Nanoparticles}
\label{nano}
  \begin{figure}[h!]
\includegraphics[scale=0.45]{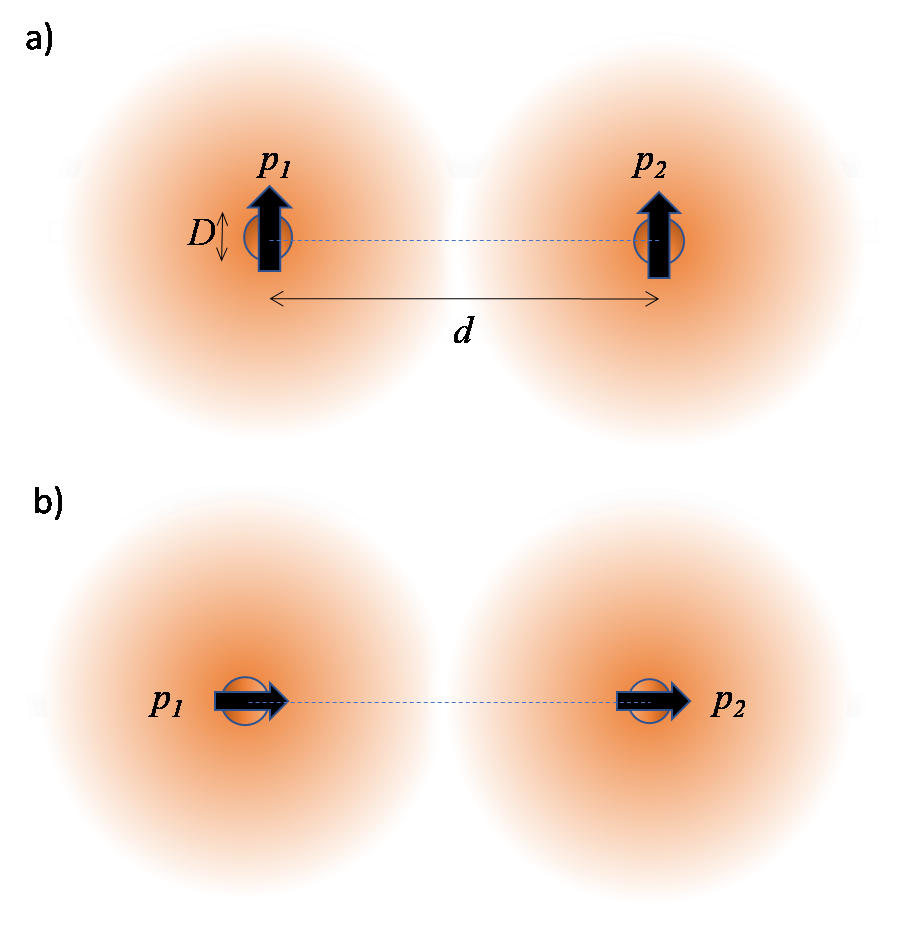}
\caption{Coupling between the fields due to the surface modes of spherical nanoparticles of diameter $D$ and center-to-center distance $d$ which is analyzed in Section \ref{nano}.  Thick arrows  indicate the direction of orientation of the dipole moments relative to the separation. The coupling strength between the fields is dependent on whether the dipoles are oriented a) side-by-side or b) head-to-tail.   }
\label{FigNP}
\end{figure}

In this section we derive the parameters of the coupled harmonic oscillator model suitable for describing the coupling of surface plasmons between two nanoparticles, and hence arrive at the expression for the heat transfer exchanged from Eq. \ref{eqHO}.
We only consider the heat transfer mediated by the coupling of surface modes and neglect contributions due to eddy currents \cite{chapuis2008radiative}, multipoles \cite{perez2008heat}, many-body effects \cite{ben2011many}. 
A similar attempt has been made to arrive at a  coupled harmonic oscillator model to 
analyze the near-field heat transfer between nanoparticles by Biehs and Agarwal \cite{biehs2013dynamical}.
The parameters in this model were arrived by comparing the steady-state heat transfer prediction 
in the coupled harmonic oscillator model with that predicted from Rytov's fluctuational 
electrodynamics theory.
Here, we will outline an alternate approach based on an analysis of splitting of eigenmodes 
to arrive at the parameters of the model.
The advantage of this approach over that followed by Biehs and Agarwal is that it can be 
used in general to analyse coupling between quasi-particles  when 
the steady state heat flux is not known before-hand.  
%

%
To find the equivalent natural frequency  $\omega_0$ and coupling constant $\gamma$  for the interaction between the surface modes across the vacuum gap  we proceed as follows:  we  use the expression for the eigenmodes of the configuration of two nanoparticles
separated by a vacuum gap using Maxwell's equations and compare this expression with that for two  coupled harmonic oscillators.
Consider first the case of a single nanoparticle with polarizability $\alpha(\omega)$. The resonant frequency is 
determined from the poles of the polarizability of the nanoparticle.  We will consider the particular case of 
spherical metallic nanoparticles with diameter $D$  whose polarizability can be expressed as:
\begin{equation}
\alpha(\omega) = \frac{\pi D^3}{2}  \frac{\varepsilon(\omega)-1}{\varepsilon(\omega)+2}
\label{alpha}
\end{equation}
When the permittivity $\varepsilon(\omega)$ of the particle is given by the Drude form:
\begin{equation}
\varepsilon(\omega) = \varepsilon_{\infty} \left(1 - \frac{\omega_L^2}{\omega (\omega + i \Gamma)} \right)
\label{dielectric}
\end{equation}
 with $\omega_p$ being the plasma frequency, and $\Gamma$ the damping parameter, the pole of Eq. \ref{alpha} gives the resonant frequency, in the limit of $\Gamma/\omega_L \ll 1$,  to be of the form $\omega = \omega_0 - i \Gamma/2 $, where $\omega_0 = \omega_L \sqrt{\varepsilon_{\infty}/(\varepsilon_{\infty} +2)}$.

 Now consider the two-particle system shown in Fig. , with their centers being separated by a gap $d$. 
 The presence of a second particle induces a change in the net polarizability of the two-particle system 
 to the form \cite{jain2006plasmon, jain2007universal}:
 \begin{equation}
 \alpha'(\omega) =  \dfrac{\alpha(\omega)}{1 - \dfrac{\kappa \alpha(\omega)}{4 \pi \varepsilon_0 d^3}}
\label{alphan}
 \end{equation}
 where $\kappa$ is an alignment factor which depends on the possible polarization modes such that  $\kappa=-1$ for dipoles aligned side-by-side (perpendicular polarization)  as shown in Fig. \ref{FigNP}a and  $\kappa=2$ for dipoles aligned head-to-tail (parallel polarization) as shown in Fig. \ref{FigNP}b.
In deriving Eq. \ref{alphan} it is assumed that the gap $d$ between the nanoparticles is such that  $1/d^3$ component of the electric field dominates compared to the other terms which vary as $1/d^2$ and $1/d$. 
The eigenmodes for this two-particle system, with the higher energy mode corresponding to the case with $\kappa = -1$ and the lower energy mode corresponding to the case with  $\kappa = 2$, can be solved for by
 substituting the expression for $\alpha$  in  Eq. \ref{alpha} into Eq. \ref{alphan} and solving for the poles of Eq. \ref{alphan}. In the limit of dipole approximation ($d\gg D$) and  under small losses ($\Gamma \ll \omega_L$) we get the eigenmodes to be of the form:
\begin{equation}
\omega_{\pm} = \omega_0^{\text{np}}\sqrt{1\pm \gamma^{\text{np}}} - i \frac{\Gamma}{2}
\label{eigenNP}
\end{equation}
where,
\begin{equation}
\omega_0^{\text{np}} = \omega_L \sqrt{\frac{\varepsilon_\infty}{\varepsilon_\infty+2}}
\label{omeganp}
\end{equation}
 and
%
\begin{equation}
\gamma^{\text{np}}(d) = \frac{9}{16} \frac{D^3}{d^3} \frac{1}{\varepsilon_\infty+2}
\label{gammanp}
\end{equation}
For large gaps $d\gg D$,  $\omega_0^{\text{np}} \approx \omega_0 $ as expected.  Comparing Eq. \ref{eigenNP} with Eq. \ref{eqeigenHO} we can directly relate the parameters of the coupled harmonic oscillator model with that of the coupled nanoparticles as follows:
\begin{equation}
\omega_0 \rightarrow \omega_0^{\text{np}}; \, \, \gamma \rightarrow \gamma^{\text{np}}; \,\,\, \delta \rightarrow \frac{\Gamma}{2}
\label{parametersNP}
\end{equation} 
 Substituting these parameters from Eq.  \ref{parametersNP} into Eq. \ref{eqHO}, we arrive at the the expression for heat transfer between two dipolar nanoparticles as:
\begin{equation}
P^{\text{np}}(d) =  \frac{\hbar \omega^{\text{np}}_0}{e^{\hbar \omega^{\text{np}}_0/(k_B T) }-1}   \left( \dfrac{  (\omega_0^{\text{np}})^2  \gamma^2 }{ (\omega_0^{\text{np}})^2  \gamma^2 +  \Gamma^2/4} \right) \frac{\Gamma}{2} 
\label{Pnp1}
\end{equation}
 For analytical simplicity if we consider the case where $k_B T \gg \hbar \omega_0$, the expression in Eq. \ref{Pnp1} can be shown to reduce to:
 \begin{equation}
P^{\text{np}}(d) =\frac{81}{128} k_B T \left( \frac{D}{d}\right)^6   \frac{\varepsilon_\infty}{(\varepsilon_\infty+2)^3}  \frac{\omega_L^2}{\Gamma}
\label{Pnp2}
\end{equation}
This expression can be compared with the expression for heat transfer between two dipolar nanoparticles derived using fluctuational electrodynamics principles \cite{volokitin2001radiative, domingues2005heat, chapuis2008radiative, perez2008heat, dedkov2010radiative}.  This, for small gaps $d/\lambda \ll 1$ reads \cite{domingues2005heat, chapuis2008radiative}:
\begin{equation}
P^{\text{np}}_{\text{FE}}(d) = \frac{3}{4 \pi^3 d^6} \int_0^\infty \frac{ \hbar \omega}{e^{\hbar \omega/(k_B T) }-1} \text{Im} \big[\alpha(\omega)\big]^2 \,\, d\omega
\label{PnpFEa}
\end{equation}
Considering the limit $k_B T \gg \hbar \omega$ and using the expression for $\alpha(\omega)$ from Eq. \ref{alpha} to evaluate the integral in Eq. \ref{PnpFEa} we get:
\begin{equation}
P^{\text{np}}_{\text{FE}}(d) =\frac{27}{128} k_B T \left( \frac{D}{d}\right)^6   \frac{\varepsilon_\infty}{(\varepsilon_\infty+2)^3}  \frac{\omega_L^2}{\Gamma}
\label{PnpFE}
\end{equation}
The expression for the heat transfer in Eq. \ref{Pnp2} derived using the coupled harmonic oscillator model  agrees with that derived using fluctuational electrodynamics principles in Eq. \ref{PnpFE} except for a numerical factor. There is currently no consensus in literature regarding the  constant numerical factor in Eq. \ref{PnpFE}  - it being 27/128 in Ref. \cite{domingues2005heat, chapuis2008radiative},  27/(32$\pi$) in Ref. \cite{volokitin2001radiative}, 27/256  in Ref. \cite{perez2008heat} and 27/8 in Ref. \cite{dedkov2010radiative}.

\section{Planar surfaces}
\label{planar}

  \begin{figure}[h!]
\includegraphics[scale=0.45]{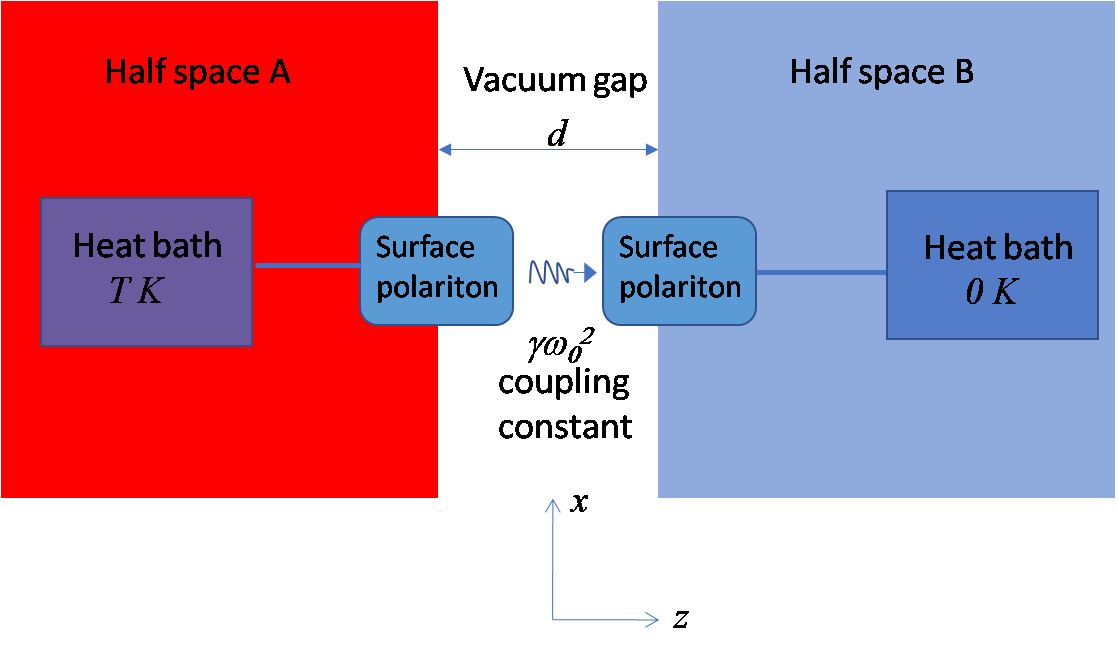}
\caption{Coupling between two surface modes located at the interface between a planar surface and vacuum which is analysed in Section \ref{planar} . The dielectric properties is taken to vary only with the frequency $\omega$ and the coupling constant $\gamma \omega_0^2$ between the surface modes is derived using Maxwell's equations}
\label{FigPl}
\end{figure}

In this section we extend the procedure outlined in Sec. \ref{nano} for deriving the expression for near-field heat transfer between two nanoparticles to that between two planar surfaces. We follow the derivation shown in Ref. \cite{barton2015classical, sasihithludynamic} to derive the equivalent parameters of the coupling oscillator, analyze the heat transfer between closely spaced metallic surfaces whose dielectric properties are given by the Drude form shown in Eq. \ref{dielectric} and contrast it with the parameters derived for coupling between nanoparticles in Sec.  \ref{nano}. %
Consider two half-spaces separated by a gap of width $d$ occupying the regions $z<-d/2$ and $z>d/2$. The in-plane two dimensional component $k_x $ and the $z$ component $k_z$ of the wavevector of a planar wave of frequency $\omega$ in the vacuum gap of this system  are related as $k_x^2+k_z^2 = (\omega/c)^2$  with $c$ being the velocity of light. From the dispersion relation for surface polaritons $\omega(k_x)$ \cite{maier2007plasmonics}  and close to the surface polariton frequency the surface mode is seen to acquire an electrostatic character with  $k_x \gg \omega/c$ (where group velocity $d\omega/d k_x \rightarrow 0$) and hence $k_z \approx ik_x$.  In this electrostatic limit the expression for the symmetric and antisymmetric forms of the potential in the vacuum gap due to the surface mode can be written as \cite{sernelius2011surface, van1968macroscopic}:
\[
\phi_{\pm}=\Phi \, \text{exp }(i k_x x - i \omega t)\begin{cases}
e^{-k_x z}+e^{k_x z}\\
e^{-k_x z}-e^{k_x z}
\end{cases}
\]
 and for $z>d/2$:
\[
\phi_{\pm}=\Phi  \, \text{exp }(ik_x x - i \omega t)\begin{cases}
(e^{-k_x d/2}+e^{k_x d/2}) & \text{exp}[-k_x (z-d/2)]\\
(e^{-k_x d/2}-e^{k_x d/2}) & \text{exp}[-k_x (z-d/2)]
\end{cases}
\]
where  $\Phi$ is an arbitrary constant. Such a form of the potential, which oscillates harmonically in time with frequency $\omega$, ensures continuity of potential at the interfaces. Satisfying continuity of perpendicular component of displacement
field gives the condition for surface modes as: 
\begin{eqnarray}
\varepsilon(\omega\pm)=\begin{cases}
\left(1-e^{k_x d}\right)/\left(1+e^{k_x d}\right)\\
\left(1+e^{k_x d}\right)/ \left(1-e^{k_x d}\right)
\end{cases}
\label{condition}
\end{eqnarray}
where
$\varepsilon(\omega)$ is the dielectric function of the half-space. Using the Drude model for the dielectric function shown in Eq. \ref{dielectric}, 
and solving for $\omega$ in Eq. \ref{condition} in the low-loss limit $\Gamma \ll \omega_L$  we obtain the eigenfrequencies of the coupled surface modes to be of the form:
 \begin{equation}
 \omega_\pm =\omega_0^{\text{pl}}( k_x, d) \sqrt{ 1 \pm \gamma^{\text{pl}}(k_x, d)  } - i \frac{\Gamma}{2}
 \label{eigensurface}
  \end{equation}
  where,
  \begin{equation}
 \omega^{\text{pl}}_0(k_x d) =\sqrt{ \varepsilon_\infty}  \omega_L \sqrt{\dfrac{  (\varepsilon_\infty + 1) - (\varepsilon_\infty - 1) e^{-2 k_x d}  }{(\varepsilon_\infty + 1)^2 - (\varepsilon_\infty - 1)^2 e^{- 2 k_x d}}}
 \label{omegap}
 \end{equation}
 and
 \begin{equation}
 \gamma^{\text{pl}}(k_x d) = \dfrac{2 e^{-k_x d} }{ (\varepsilon_\infty + 1) - (\varepsilon_\infty - 1) e^{-2 k_x d}  } ;
 \label{xip}
 \end{equation}
In the limit of large gaps i.e., $k_x  d \rightarrow \infty$ we have  $\omega_0^{\text{pl}}(k_x d) \rightarrow \omega_L \sqrt{\varepsilon_\infty/(\varepsilon_\infty+1)}$ which is the surface plasmon frequency for a single half-space, and $\gamma^{\text{pl}}(k_x d) \rightarrow 0$ as expected.  From Eq. \ref{eqeigenHO} and \ref{eigensurface} we have for planar surfaces:
\begin{equation}
\omega_0 \rightarrow \omega_0^{\text{pl}}(k_x d); \, \, \gamma \rightarrow \gamma^{\text{pl}}(k_x d); \,\,\, \delta \rightarrow \frac{\Gamma}{2}
\label{parametersPl}
\end{equation} 

Comparing the coupling parameters  $\omega^{\text{pl}}_0$ and  $\gamma^{\text{pl}}$  derived in Eqs. \ref{omegap} and  Eq. \ref{xip} for planar surfaces with those derived for nanoparticles  $\omega^{\text{np}}_0$ and  $\gamma^{\text{np}}$  in Eqs. \ref{omeganp} and Eq. \ref{gammanp} it is observed that in the former case the coupling parameters are a function of the in-plane wavevector modes $k_x$. We thus have a spectrum of oscillators  (labelled by the modes $k_x$) and since only modes of the same in-plane wavevector components interact across the vacuum gap we can consider each of the oscillators to contribute independently to the heat transfer. 
We can thus write the expression for the near-field heat transfer between planar surfaces from Eq. \ref{eqHO} and Eq. \ref{parametersPl} to be of the form:
\begin{equation}
P^{\text{pl}}(d) = \frac{1}{4 \pi^2} \int \frac{\hbar \omega^{\text{pl}}_0 (k_x d)}{e^{\hbar \omega^{\text{pl}}_0(k_x d)/(k_B T) }-1}   \left( \dfrac{  [\omega_0^{\text{pl}}(k_x d)]^2  \gamma^2(k_x d) }{ [\omega_0^{\text{pl}}(k_x d)]^2  \gamma^2(k_x d) +  \Gamma^2/4} \right) \frac{\Gamma}{2}  \,\, d^2 k_x
\label{Ppl1}
\end{equation}
For analytical simplicity considering the case when $k_B T \gg \hbar \omega^{\text{pl}}_0$ and substituting the expressions for the coupling parameters from Eqs. \ref{omegap} and \ref{xip},  Eq. \ref{Ppl1} reduces to:
\begin{equation}
P^{\text{pl}}(d) = \frac{k_B T}{2 \pi d^2} \int_0^\infty   \dfrac{8 \varepsilon_\infty \omega_L^2  }{e^{- 2 x} \Gamma( \varepsilon_\infty - 1)^3 + \Gamma e^{ 2 x} ( \varepsilon_\infty + 1)^3 - 2 \Gamma  \varepsilon_\infty ( \varepsilon_\infty^2 - 1) + 16  \varepsilon_\infty \omega_L^2 \Gamma^{-1}  }    \,\,  x \, d x
\label{Ppl2}
\end{equation}
This expression can be compared with the expression of near-field heat transfer between two planar surfaces derived from  fluctuational electrodynamics principles \cite{loomis94} which, in the limit $k_x d \rightarrow 0$ and $k_B T \gg \hbar \omega $, reads :
\begin{equation}
P^{\text{pl}}_{\text{FE}}(d)= \frac{k_B T}{\pi^2 d^2} \int_0^\infty  \, d\omega \,  \int_0^\infty \frac{ \left[\text{Im} \varepsilon(\omega)\right]^2 x e^{-x}  \, dx}{|(\varepsilon(\omega)+1)^2 - (\varepsilon(\omega) -1)^2 e^{-x}|^2}
\label{QFD}
\end{equation}
Substituting the expression for the dielectric function $\varepsilon(\omega)$  from Eq. \ref{dielectric} and expanding,  Eq. \ref{QFD} reduces to the form:
\begin{equation}
P^{\text{pl}}_{\text{FE}}(d)=  \frac{k_B T}{\pi^2 d^2} \int_0^\infty x e^{-x} \, dx \int_0^\infty \frac{M(\omega)}{N(\omega)} \, d\omega
\label{cauchy}
\end{equation}
where $M(\omega)$ and $N(\omega) $ are complicated polynomial even functions of $\omega$. By carrying out the integral over $\omega$ in the complex plane using Cauchy's residue theorem  the expression in Eq. \ref{cauchy} reduces to that in Eq. \ref{Ppl2}. 

To conclude, we have arrived at an expression for heat transfer between two heat baths maintained by constant temperatures mediated by a coupled harmonic oscillator system using Green's function theory and shown its equivalence to those derived in literature using Master equation and Langevin dynamics.  We use this expression to find surface plasmon mediated heat transfer between two closely spaced nanoparticles, and also for two closely spaced planar surfaces whose dielectric properties are of the Drude form.  In order to establish an equivalence between the coupled harmonic oscillator system and the coupled nanoparticles/planar surfaces configuration we have compared the splitting in the eigenmodes of the two systems to arrive at the equivalent coupling parameters.  The expression of heat transfer thus obtained for both these configurations is shown to be consistent with that obtained using the established theory of fluctuational electrodynamics.

\label{section6}


\section*{Acknowledgements}
We acknowledge useful discussions with Prof. Girish Sarin Agarwal and with Dr. Svend-Age Biehs.
\section*{References}


\bibliographystyle{ieeetr}

\begin{thebibliography}{10}

\bibitem{narayanaswamy03a}
A.~Narayanaswamy and G.~Chen, ``Surface modes for near field
  thermophotovoltaics,'' {\em Appl. Phys. Lett.}, vol.~82, pp.~3544--3546,
  2003.

\bibitem{dimatteo2001enhanced}
R.~S. DiMatteo, P.~Greiff, S.~L. Finberg, K.~A. Young-Waithe, H.~Choy, M.~M.
  Masaki, and C.~G. Fonstad, ``Enhanced photogeneration of carriers in a
  semiconductor via coupling across a nonisothermal nanoscale vacuum gap,''
  {\em Applied Physics Letters}, vol.~79, no.~12, pp.~1894--1896, 2001.

\bibitem{laroche06b}
M.~Laroche, R.~Carminati, and J.-J. Greffet, ``Near-field thermophotovoltaic
  energy conversion,'' {\em J. of Appl. Phys.}, vol.~100, p.~063704, 2006.

\bibitem{otey2010thermal}
C.~R. Otey, W.~T. Lau, and S.~Fan, ``Thermal rectification through vacuum,''
  {\em Phys. Rev. Lett}, vol.~104, no.~15, p.~154301, 2010.

\bibitem{rytov59a}
S.~M. Rytov, {\em Theory of Electric Fluctuations and Thermal Radiation}.
\newblock Bedford, MA: Air Force Cambridge Research Center, 1959.

\bibitem{polder71}
D.~Polder and M.~Van~Hove, ``Theory of radiative heat transfer between closely
  spaced bodies,'' {\em Phys. Rev. B}, vol.~4, pp.~3303--3314, 1971.

\bibitem{loomis94}
J.~J. Loomis and H.~J. Maris, ``Theory of heat transfer by evanescent
  electromagnetic waves,'' {\em Phs. Rev. B}, vol.~50, p.~18517, 1994.

\bibitem{pendry1999radiative}
J.~Pendry, ``{Radiative exchange of heat between nanostructures},'' {\em
  Journal of Physics: Condensed Matter}, vol.~11, p.~6621, 1999.

\bibitem{joulain05a}
K.~Joulain, J.-P. Mulet, F.~Marquier, R.~Carminati, and J.-J. Greffet,
  ``{Surface electromagnetic waves thermally excited: radiative heat transfer,
  coherence properties and {C}asimir forces revisited in the near field},''
  {\em Surf. Sci. Rep.}, vol.~57, pp.~59 -- 112, 2005.

\bibitem{song2016radiative}
B.~Song, D.~Thompson, A.~Fiorino, Y.~Ganjeh, P.~Reddy, and E.~Meyhofer,
  ``Radiative heat conductances between dielectric and metallic parallel plates
  with nanoscale gaps,'' {\em Nature nanotechnology}, vol.~11, no.~6, p.~509,
  2016.

\bibitem{kim2015radiative}
K.~Kim, B.~Song, V.~Fern{\'a}ndez-Hurtado, W.~Lee, W.~Jeong, L.~Cui,
  D.~Thompson, J.~Feist, M.~T. Reid, F.~J. Garc{\'\i}a-Vidal, {\em et~al.},
  ``Radiative heat transfer in the extreme near field,'' {\em Nature},
  vol.~528, no.~7582, p.~387, 2015.

\bibitem{kloppstech2017giant}
K.~Kloppstech, N.~K{\"o}nne, S.-A. Biehs, A.~W. Rodriguez, L.~Worbes,
  D.~Hellmann, and A.~Kittel, ``Giant heat transfer in the crossover regime
  between conduction and radiation,'' {\em Nature Communications}, vol.~8,
  no.~14475, 2017.

\bibitem{rousseau2009radiative}
E.~Rousseau, A.~Siria, G.~Jourdan, S.~Volz, F.~Comin, J.~Chevrier, and J.-J.
  Greffet, ``Radiative heat transfer at the nanoscale,'' {\em Nature
  Photonics}, vol.~3, no.~9, pp.~514--517, 2009.

\bibitem{shen2009surface}
S.~Shen, A.~Narayanaswamy, and G.~Chen, ``Surface phonon polaritons mediated
  energy transfer between nanoscale gaps,'' {\em Nano Letters}, vol.~9, no.~8,
  pp.~2909--2913, 2009.

\bibitem{biehs2013dynamical}
S.-A. Biehs and G.~S. Agarwal, ``Dynamical quantum theory of heat transfer
  between plasmonic nanosystems,'' {\em JOSA B}, vol.~30, no.~3, pp.~700--707,
  2013.

\bibitem{barton2015classical}
G.~Barton, ``Classical van der waals heat flow between oscillators and between
  half-spaces,'' {\em Journal of Physics: Condensed Matter}, vol.~27, no.~21,
  p.~214005, 2015.

\bibitem{sasihithludynamic}
K.~Sasihithlu and G.~S. Agarwal, ``Dynamic near-field heat transfer between
  macroscopic bodies for nanometric gaps,'' {\em Nanophotonics}.

\bibitem{chance1978molecular}
R.~Chance, A.~Prock, and R.~Silbey, ``Molecular fluorescence and energy
  transfer near interfaces,'' {\em Adv. Chem. Phys}, vol.~37, no.~1, p.~65,
  1978.

\bibitem{le2008principles}
E.~Le~Ru and P.~Etchegoin, {\em Principles of Surface-Enhanced Raman
  Spectroscopy: and related plasmonic effects}.
\newblock Elsevier, 2008.

\bibitem{dorofeyev2013coupled}
I.~Dorofeyev, ``Coupled quantum oscillators within independent quantum
  reservoirs,'' {\em Canadian Journal of Physics}, vol.~91, no.~7,
  pp.~537--541, 2013.

\bibitem{ghesquiere2013entanglement}
A.~Ghesqui{\`e}re and T.~Dorlas, ``Entanglement of a two-particle gaussian
  state interacting with a heat bath,'' {\em Physics Letters A}, vol.~377,
  no.~40, pp.~2831--2839, 2013.

\bibitem{pendry2016phonon}
J.~Pendry, K.~Sasihithlu, and R.~Craster, ``Phonon-assisted heat transfer
  between vacuum-separated surfaces,'' {\em Physical Review B}, vol.~94, no.~7,
  p.~075414, 2016.

\bibitem{reif2009fundamentals}
F.~Reif, {\em Fundamentals of statistical and thermal physics}.
\newblock Waveland Press, 2009.

\bibitem{chapuis2008radiative}
P.-O. Chapuis, M.~Laroche, S.~Volz, and J.-J. Greffet, ``Radiative heat
  transfer between metallic nanoparticles,'' {\em Applied Physics Letters},
  vol.~92, no.~20, p.~201906, 2008.

\bibitem{perez2008heat}
A.~P{\'e}rez-Madrid, J.~M. Rub{\'\i}, and L.~C. Lapas, ``Heat transfer between
  nanoparticles: Thermal conductance for near-field interactions,'' {\em
  Physical Review B}, vol.~77, no.~15, p.~155417, 2008.

\bibitem{ben2011many}
P.~Ben-Abdallah, S.-A. Biehs, and K.~Joulain, ``Many-body radiative heat
  transfer theory,'' {\em Physical Review Letters}, vol.~107, no.~11,
  p.~114301, 2011.

\bibitem{jain2006plasmon}
P.~K. Jain, S.~Eustis, and M.~A. El-Sayed, ``Plasmon coupling in nanorod
  assemblies: optical absorption, discrete dipole approximation simulation, and
  exciton-coupling model,'' {\em The Journal of Physical Chemistry B},
  vol.~110, no.~37, pp.~18243--18253, 2006.

\bibitem{jain2007universal}
P.~K. Jain, W.~Huang, and M.~A. El-Sayed, ``On the universal scaling behavior
  of the distance decay of plasmon coupling in metal nanoparticle pairs: a
  plasmon ruler equation,'' {\em Nano Letters}, vol.~7, no.~7, pp.~2080--2088,
  2007.

\bibitem{volokitin2001radiative}
A.~Volokitin and B.~Persson, ``Radiative heat transfer between
  nanostructures,'' {\em Physical Review B}, vol.~63, no.~20, p.~205404, 2001.

\bibitem{domingues2005heat}
G.~Domingues, S.~Volz, K.~Joulain, and J.-J. Greffet, ``Heat transfer between
  two nanoparticles through near field interaction,'' {\em Physical review
  letters}, vol.~94, no.~8, p.~085901, 2005.

\bibitem{dedkov2010radiative}
G.~Dedkov and A.~Kyasov, ``Radiative heat transfer of spherical particles
  mediated by fluctuation electromagnetic field,'' {\em Journal of
  Computational and Theoretical Nanoscience}, vol.~7, no.~10, pp.~2019--2023,
  2010.

\bibitem{maier2007plasmonics}
S.~A. Maier, {\em Plasmonics: fundamentals and applications}.
\newblock Springer Science \& Business Media, 2007.

\bibitem{sernelius2011surface}
B.~E. Sernelius, {\em Surface modes in physics}.
\newblock Wiley-Vch, 2011.

\bibitem{van1968macroscopic}
N.~Van~Kampen, B.~Nijboer, and K.~Schram, ``On the macroscopic theory of van
  der waals forces,'' {\em Physics letters A}, vol.~26, no.~7, pp.~307--308,
  1968.

\end{thebibliography}

\end{document}